\documentclass[preprint,12pt]{elsarticle}

\usepackage{amssymb}
\usepackage{amsmath}
\usepackage{subfigure}
\begin{document}

\begin{frontmatter}



\title{Representation for the Pyrochlore Lattice}


\author[label1]{Andr\'e Luis Passos}
\address[label1]{DFI, CCET, Universidade Federal de Sergipe, S\~ao Crist\'ov\~ao, SE, Brazil}
\ead{passosal@gmail.com}

\author[label2]{Douglas F. de Albuquerque}
\address[label2]{DMA, CCET, Universidade Federal de Sergipe, S\~ao Crist\'ov\~ao, SE, Brazil}

\author[label3]{Jo\~ao Batista Santos Filho}
\address[label3]{Instituto Federal de Sergipe, S\~ao Crist\'ov\~ao, SE, Brazil}

\begin{abstract}

This work aims to present a representation for the Kagom\'e and pyrochlore lattices for Monte Carlo simulation and some results of the critical properties.
These geometric frustrated lattices are composed of corner sharing triangles and tetrahedrons respectively.
The simulation was performed with the Cluster Wollf Algorithm for the spins updates using the standard ferromagnetic Ising Model.
The determination of the critical temperature and exponents was done through the Histogram Technique and the Finite-Size Scalling Theory. 

\end{abstract}


\end{frontmatter}



\section{Introduction}
\label{sec:Intro}

For geometric frustrated magnetic materials it is impossible to choose one conguration with minimal energy by the individual minimization of the magnetic moments interactions.
Consequently, their ground states are degenerated.
The most studied cases are systems with crystalline structures composed by triangles and tetrahedrons with nearest-neighbor antiferromagnetic interaction of their magnetic moments~\cite{Melko2004}.
Recently frustration has been observed in jarosites and pyrochlore oxides as a result of ferromagnetic interaction~\cite{Castelnovo2011}.
In these structures the spins are arranged in Kagom\'e and pyrochlore lattices and forced to point in towards the center of the triangles and tetrahedrons respectively.
For this reason, there is only the two possible states ,``in'' or ``out'', and therefore can be approximated through the classical Ising model.
Since the angle between the spins the ferro-antiferromagnetic rule is changed so ferromagnetic systems can be mapped by the antiferromagnetic Ising model and vice-versa~\cite{Bramwell1998}.

These systems are called spin ice because of the similarity between their ground state conguration and the Pauli model for the water ice.
Due to the simplicity of the Ising model in comparison with Heisenberg they became a model for the study of geometric frustration.
In some cases it is even possible an analytic treatment in order to compare to experimental and computational results~\cite{Garcia-Adeva2001}.
\begin{figure}[!ht]
  \centering
  \includegraphics[angle=90,scale=0.27]{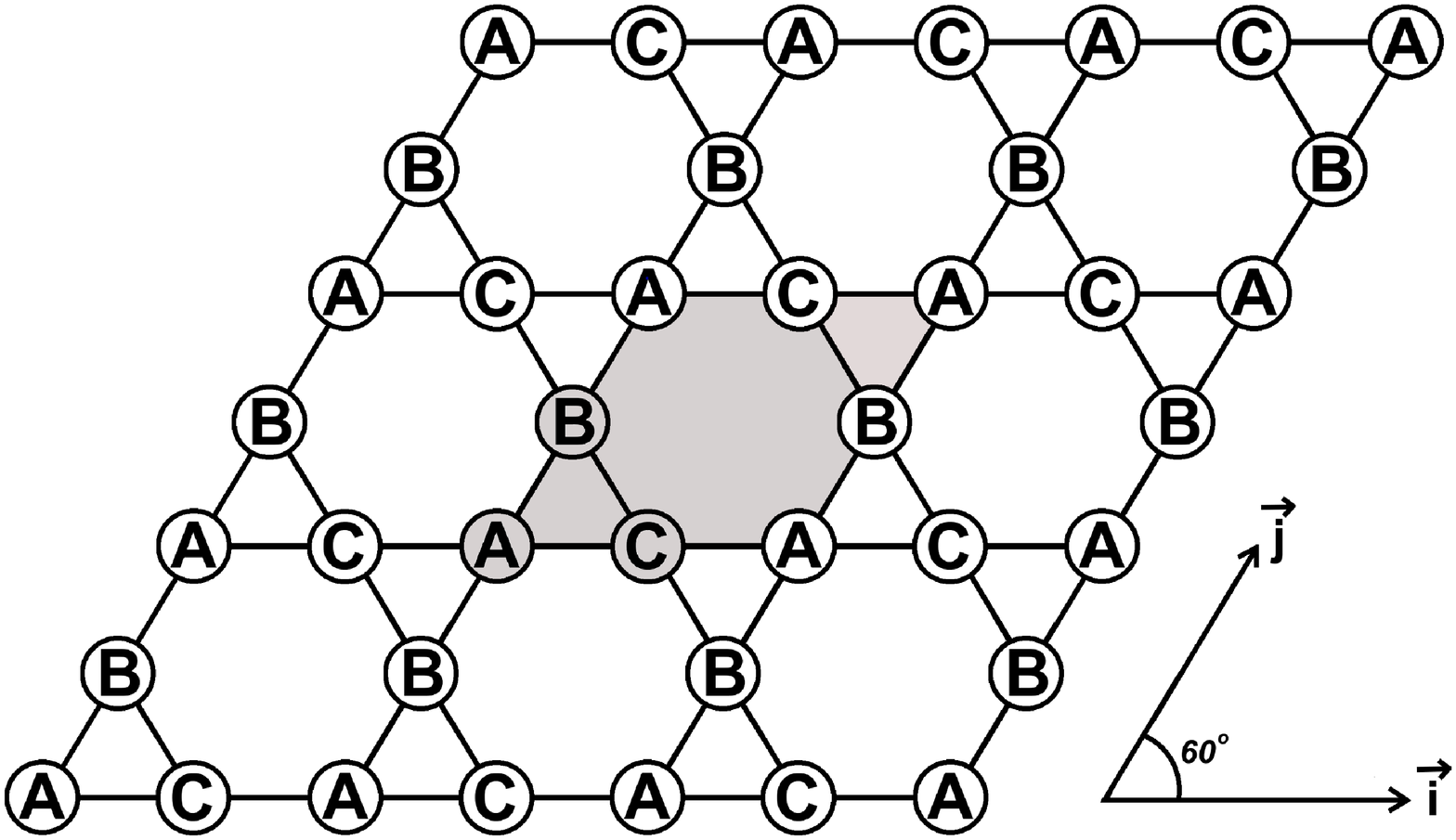}
  \caption{Representation of the sites in the Kagom\'e lattice.}\label{fig:Kagome}
\end{figure}

The recent artificial production of the first spin ices in the square and honeycomb lattices had attracted a lot of attention~\cite{Li2013}.
They are constructed with bi-dimensional sequences of nanoislands with ferromagnetic magnetic moments considered as effective Ising spins.
Recently it was possible to produce artificial jarosite and volborthite crystals with practically no distortions or impurities, something not possible for natural materials found in nature~\cite{Grohol2005,Matan2011}.
Another recent example of the Ising model is the study of magnetic materials in the triangular Kagom\'e lattice, artificially produced, as an alternative for adiabatic demagnetization~\cite{Pecharsky1999}.
They consist of a Kagom\'e lattice with an inserted site between two nearest-neighbors.
Additionally it can be used in the study of the exchange coupling constants~\cite{Loh2008}.
They are composed of sequences Kagom\'e lattices with weak interaction between the layers so they are considered bidimensional structures.

In this paper it is presented an alternative to the conventional representation of the Kagom\'e~\cite{Ghaemi2012} and pyrochlore~\cite{Lin2011} lattices for Monte Carlo study of spins systems.
Furthermore, the critical temperatures and exponents using the standard ferromagnetic Ising model is shown.
As suggested, there are the studies of the XY and Heisenberg models for these lattices or the addition of the dipolar term in the Ising model~\cite{Chern2011}.
There is also the possibility of the study of critical quantities and phase diagram for diluted cases.


\section{Method}
\label{sec:Methods}

The Kagom\'e lattice is a two-dimensional arrangement of sites with four nearest-neighbors as shown in Figure~\ref{fig:Kagome}.
The unitary cell in gray has three sites with bonds forming equilateral triangles, each site in an  indenpendent rhombic sublattices labeled A, B and C.
For a site with label $t$ and coordinates $i$ and $j$ its nearest-neighbors are defined using the two versors \textbf{\^{i}} and \textbf{\^{j}} as shown in Table~\ref{tab:Kag_neighbors} ($t = 0$, $1$ and $2$ for A, B and C respectivelly).
The sites are indexed by $i + \left( j-1 \right)L + tL^{2}$, the space between the sites are set to 1 and $\left\{ i,j \in \mathbb{N}^{*} | i,j < L\right\} $.

\begin{figure}
\begin{minipage}[t]{0.5\textwidth}\subfigure[]{
\includegraphics[angle=90,width=\textwidth]{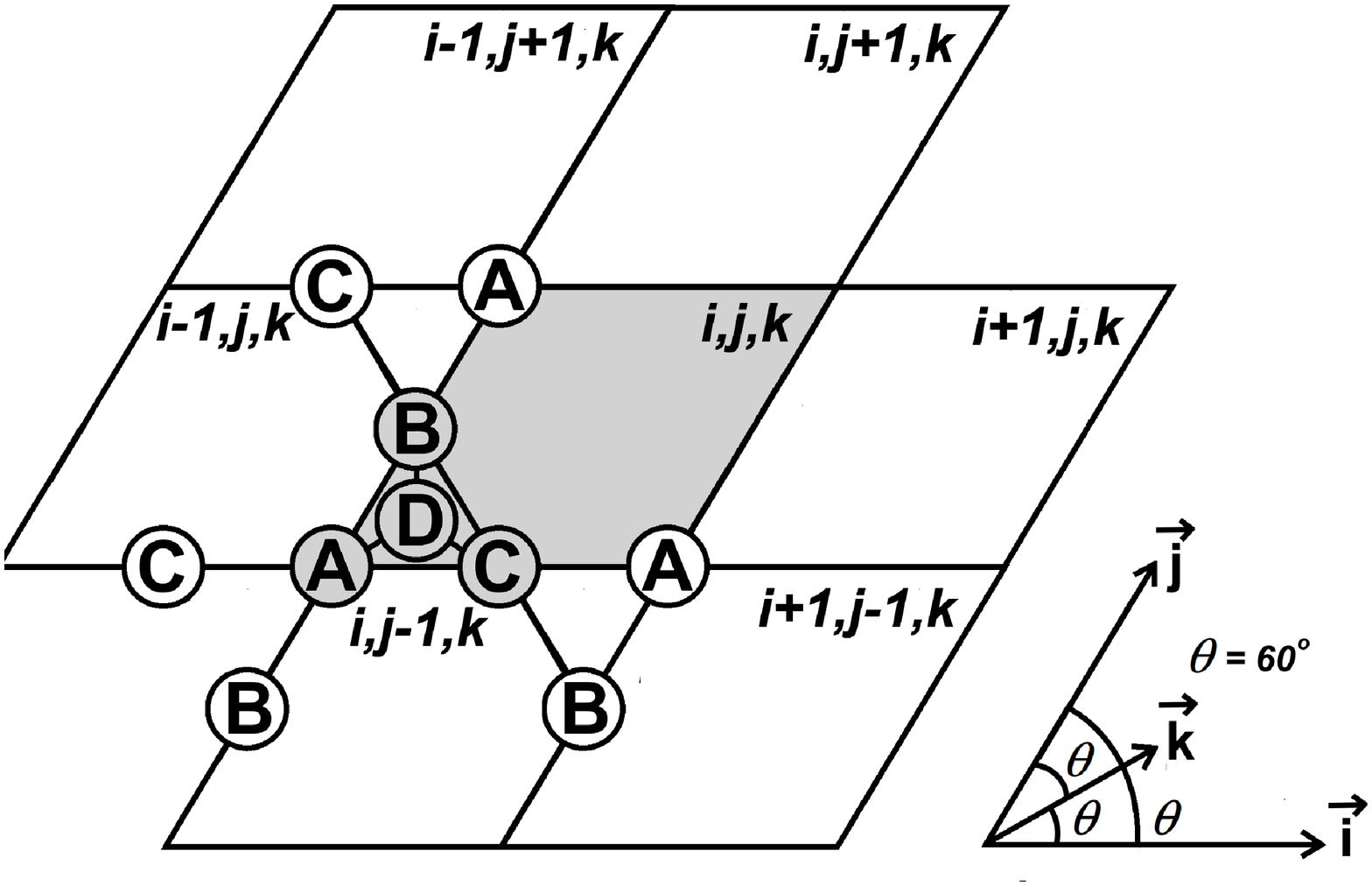}
}
\end{minipage}
\begin{minipage}[t]{0.7\textwidth}\subfigure[]{
\includegraphics[angle=90,width=\textwidth]{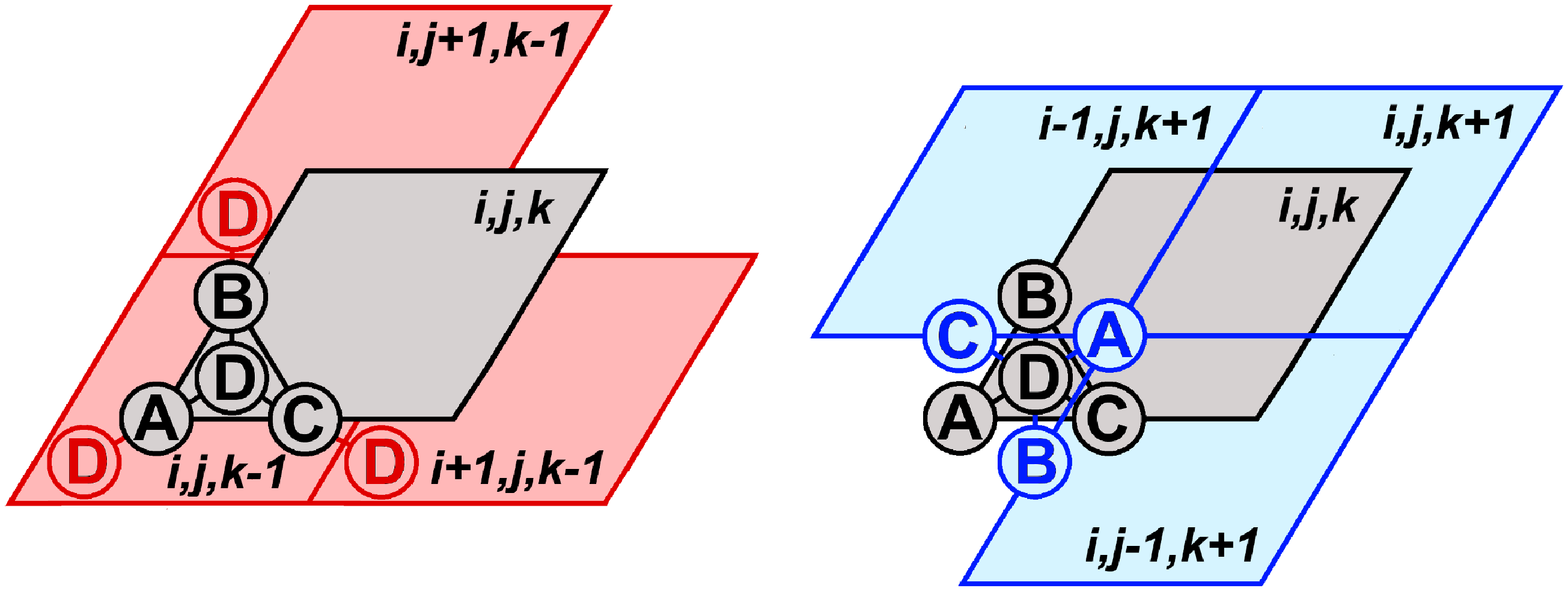}
}
\end{minipage}
\caption{Nearest neighbors for the sites of the unitary cell for the pyrochlore lattice (a) in the same plane $k$ and (b) previous and next planes $k-1$ and $k+1$ respectivelly. The $z$ component of the vector $\textbf{k}$ points out of the plane.}\label{fig:pyrochlore} 
\end{figure}

\begin{table}[h]
  \centering
  \caption{Nearest-neighbors for the Kagom\'e lattice.}
  \label{tab:Kag_neighbors}
  \begin{tabular}{ccc}
    \hline
    Sub-lattice A & Sub-lattice B & Sub-lattice C \\
    \hline
    $(i,j)$ B   & $(i,j)$ C     & $(i,j)$ A \\
    $(i,j-1)$ B & $(i-1,j+1)$ C & $(i+1,j)$ A \\
    $(i,j)$ C   & $(i,j)$ A     & $(i,j)$ B \\
    $(i-1,j)$ C & $(i,j+1)$ A   & $(i+1,j-1)$ B \\
    \hline
  \end{tabular}
\end{table}
On the other hand, the pyrochlore lattice is the three-dimensional analogue of the Kagom\'e with six nearest-neighbors, and it is constructed using the same idea.
A forth site labeled D ($t = 3$) is added to the Kagom\'e unitary cell, immediately above it, in a way all the sites are equally spaced and their bonds form a tetrahedra.
It is represented in Figure~\ref{fig:pyrochlore} with the upper and lower layers of the unitary cell.
The four sites form independent ortorhombic lattices and its nearest-neighbors, taking into account the versor \textbf{\^{k}}, are listed in Table~\ref{tab:pyro_neighbors}.
The sites are indexed by $i + \left( j-1 \right)L + \left( k-1 \right)L^{2} + tL^{3}$ with $\left\{ i,j,k \in \mathbb{N}^{*} | i,j,k < L\right\} $.

To reduce the finite size effects, the periodic boundary condition is used. This implies in the substitution of the term $(i+1)$ by $1$ for $i=L$ and $(i-1)$ by $L$ for $i=1$ in Tables \ref{tab:Kag_neighbors} and \ref{tab:pyro_neighbors} (same for $j$ and $k$).

It was used the zero-field standard Ising model
\begin{align}
\mathcal{H} = -J\sum_{\left< i,j \right> }\sigma_i\sigma_j,
\end{align}
for the Monte Carlo Simulation with the Cluster Wolff Algorithm for the spin updates~\cite{Landau2009} where $J$ is the exchange coupling constant, $\sigma = \pm 1$ is the spin state and $\left< i,j \right> $ indicates that the sum is only over the nearest neighbors.
It was discarded 20\% of the MCS (Monte Carlo Steps) for the termalization so a pair of energy and magnetization for the rest of the MCS is collected.
The number of MCS was chosen so that the mean values termodynamic quantities (mean energy and magnetization, specific heat, magnetic susceptibility and Binder Cumulant) do not change for larger number of steps.
The values of these quantities in the vicinity of the temperature where the simulation was performed is than obtained by the Histogram Technique~\cite{Landau2009,Binder2010}.

\begin{table}[th]
  \centering
  \caption{Nearest-neighbors for the pyrochlore lattice.}
  \label{tab:pyro_neighbors}
  \begin{tabular}{cccc}
    \hline
    Sublattice A & Sublattice B & Sublattice C & Sublattice D \\
    \hline
    $(i,j,k)$ B   & $(i,j,k)$ C     & $(i,j,k)$ D     & $(i,j,k)$ A \\
    $(i,j-1,k)$ B & $(i-1,j+1,k)$ C & $(i+1,j,k-1)$ D & $(i,j,k+1)$ A \\
    $(i,j,k)$ C   & $(i,j,k)$ D     & $(i,j,k)$ A     & $(i,j,k)$ B \\
    $(i-1,j,k)$ C & $(i,j+1,k-1)$ D & $(i+1,j,k)$ A   & $(i,j-1,k+1)$ B \\
    $(i,j,k)$ D   & $(i,j,k)$ A     & $(i,j,k)$ B     & $(i,j,k)$ C \\
    $(i,j,k-1)$ D & $(i,j+1,k)$ A   & $(i+1,j-1,k)$ B & $(i-1,j,k+1)$ C \\
    \hline
  \end{tabular}
\end{table}

According to the Finite-Size Scaling Theory~\cite{Brankov1996} the critical temperature $T_C$ can be obtained by the value for the temperature in which the Binder Cumulant ( $U_4 = 1 - \frac{\langle m^4 \rangle}{3 \langle m^2 \rangle}\,$) of different size systems intercept themselves.
Due to the scalling laws its only necessary any pair of the critical exponents to obtain the values for the remaining~\cite{Brankov1996}.

The magnetization per site $m$ and the derivative of the Binder Cumulant with respect to the temperature ${\partial U_4}/{\partial T}$ it is given by
\begin{subequations}
\begin{equation}
m \propto L^{-\beta / \nu},
\end{equation}
\begin{equation}
{\partial U_4}/{\partial T} \propto L^{1/ \nu}.
\end{equation}
\label{eq:exponents}
\end{subequations}


\begin{figure}[!h]
\begin{minipage}[t]{0.5\textwidth}\subfigure[]{
\includegraphics[width=\textwidth]{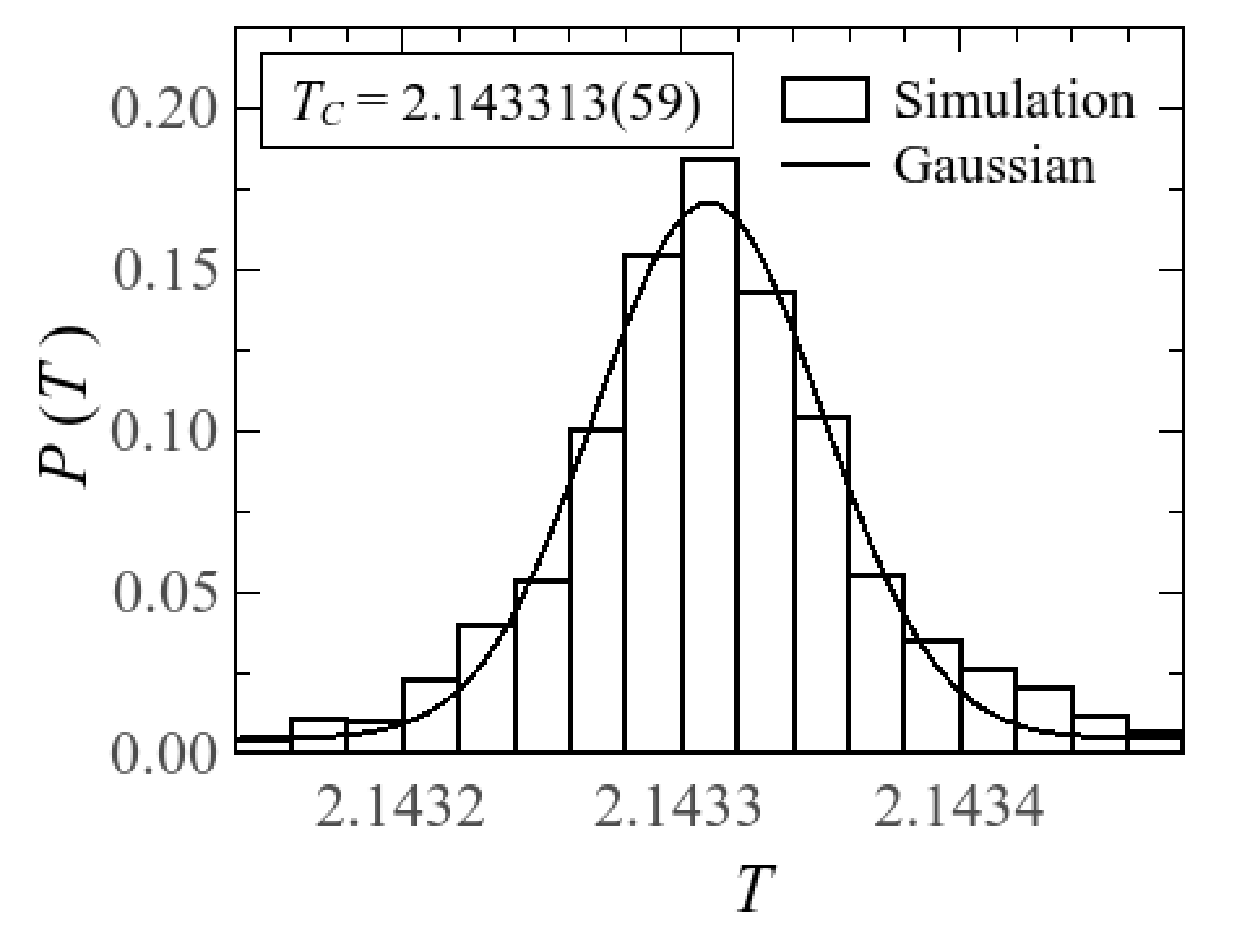}
}
\end{minipage}
\begin{minipage}[t]{0.5\textwidth}\subfigure[]{
\includegraphics[width=\textwidth]{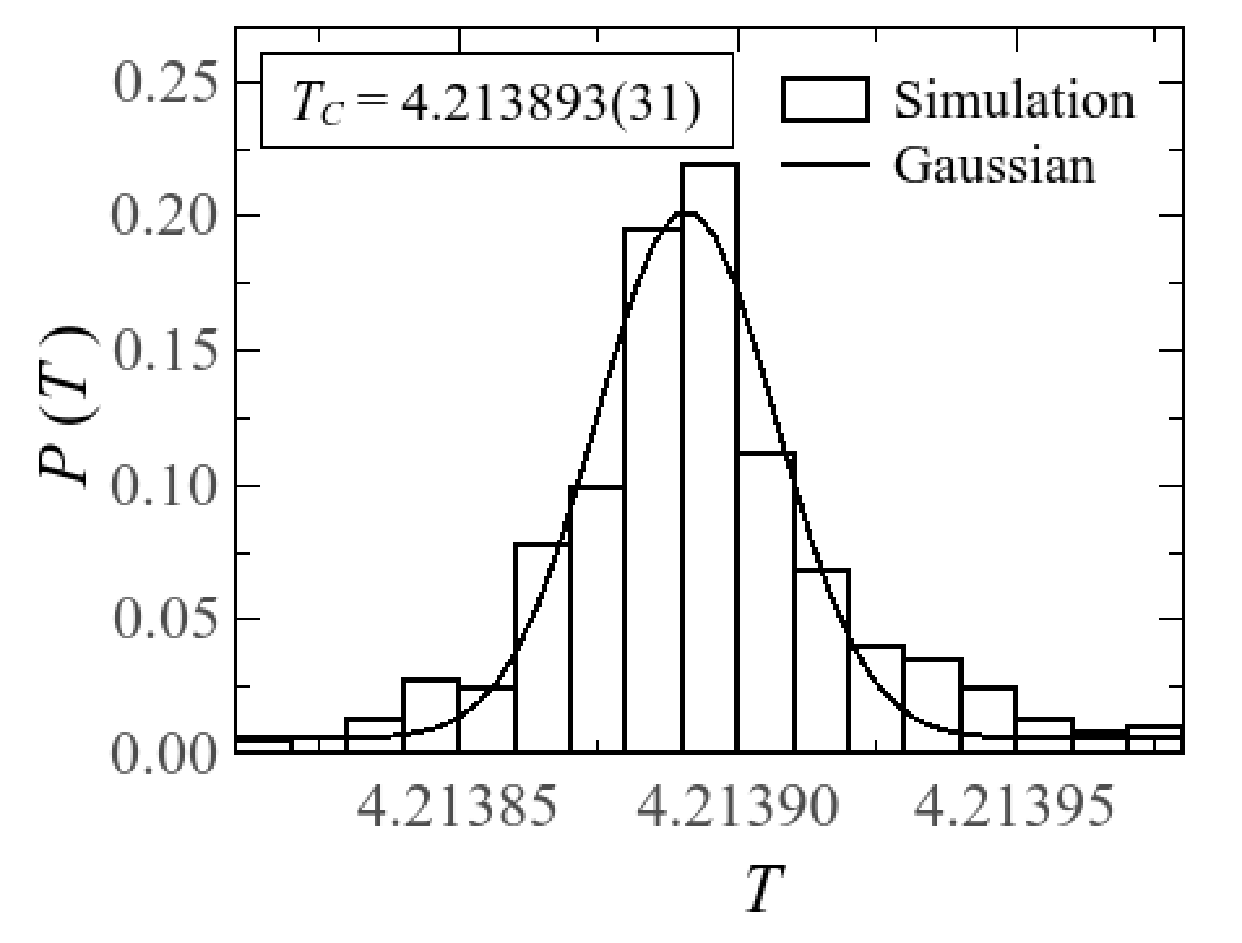}
}
\end{minipage}\caption{Graphs of the critical temperature distribution $P(T_C)$ for (a) Kagom\'e and (b) pyrochlore lattices.}\label{fig:TC}
\end{figure}
\section{Results and Remarks}
\label{sec:Results_Remarks}

\begin{figure}[!h]
\begin{minipage}[t]{0.5\textwidth}\subfigure[]{
\includegraphics[angle=0,width=\textwidth]{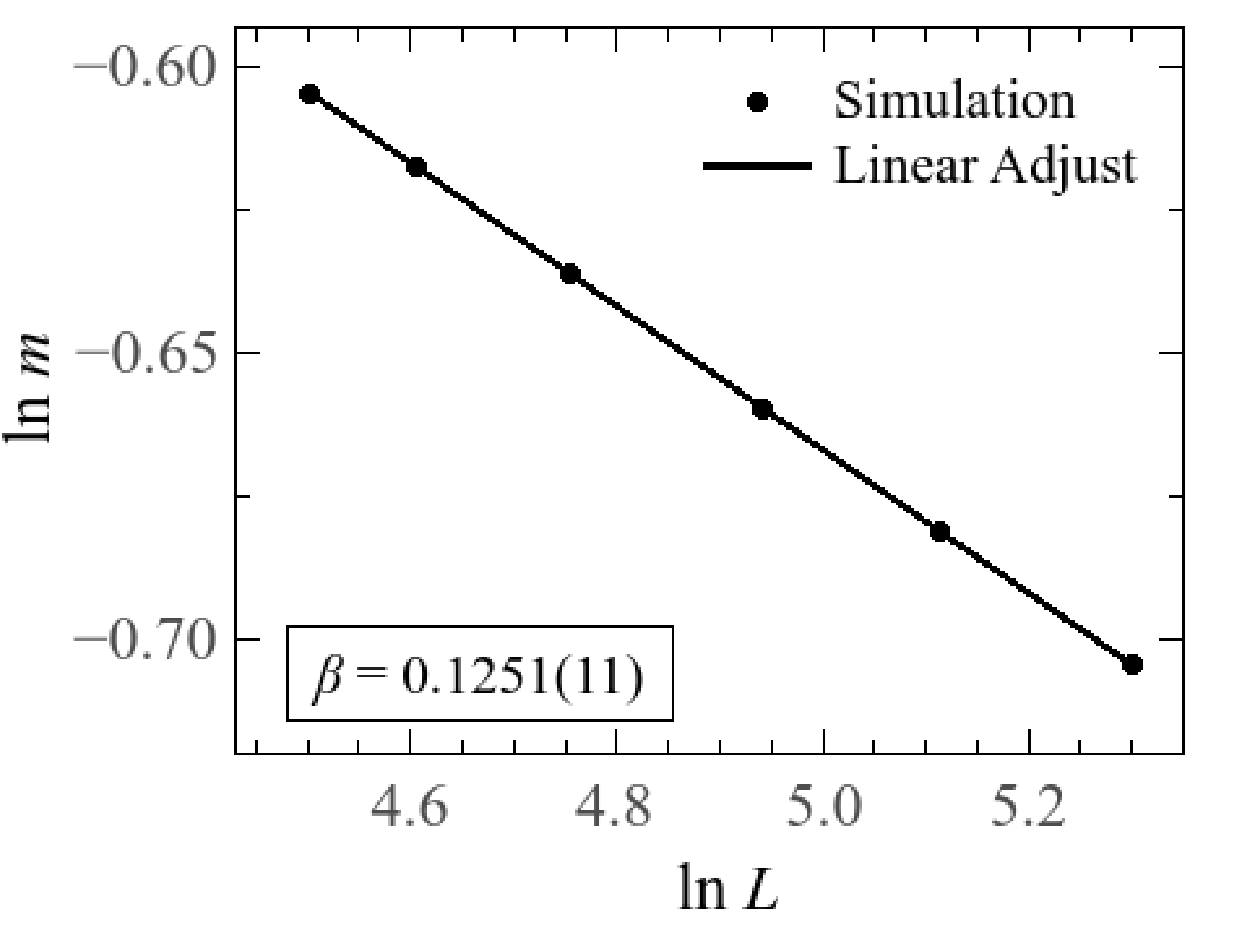}
}
\end{minipage}
\begin{minipage}[t]{0.5\textwidth}\subfigure[]{
\includegraphics[angle=0,width=\textwidth]{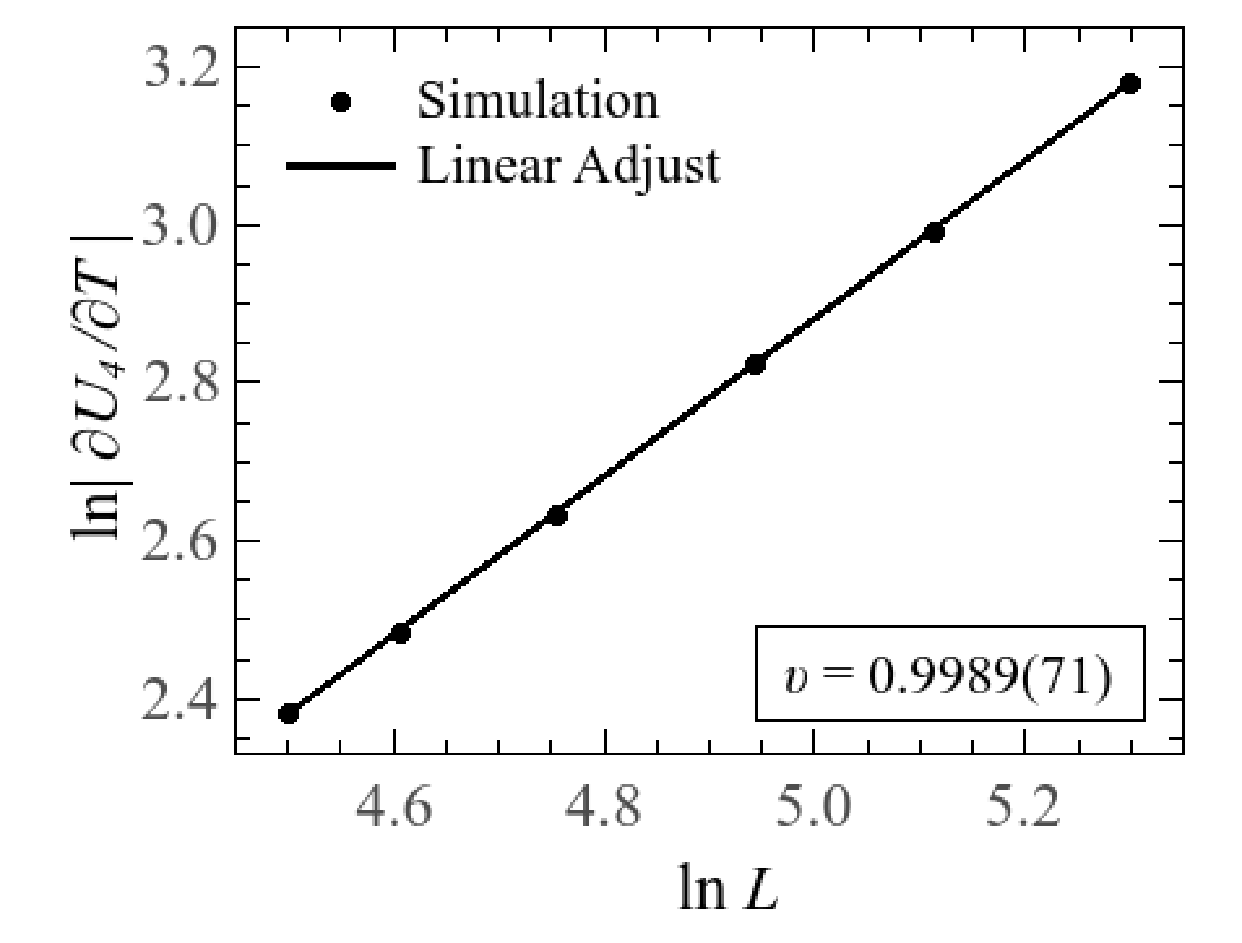}
}
\end{minipage}
\begin{minipage}[t]{0.5\textwidth}\subfigure[]{
\includegraphics[angle=0,width=\textwidth]{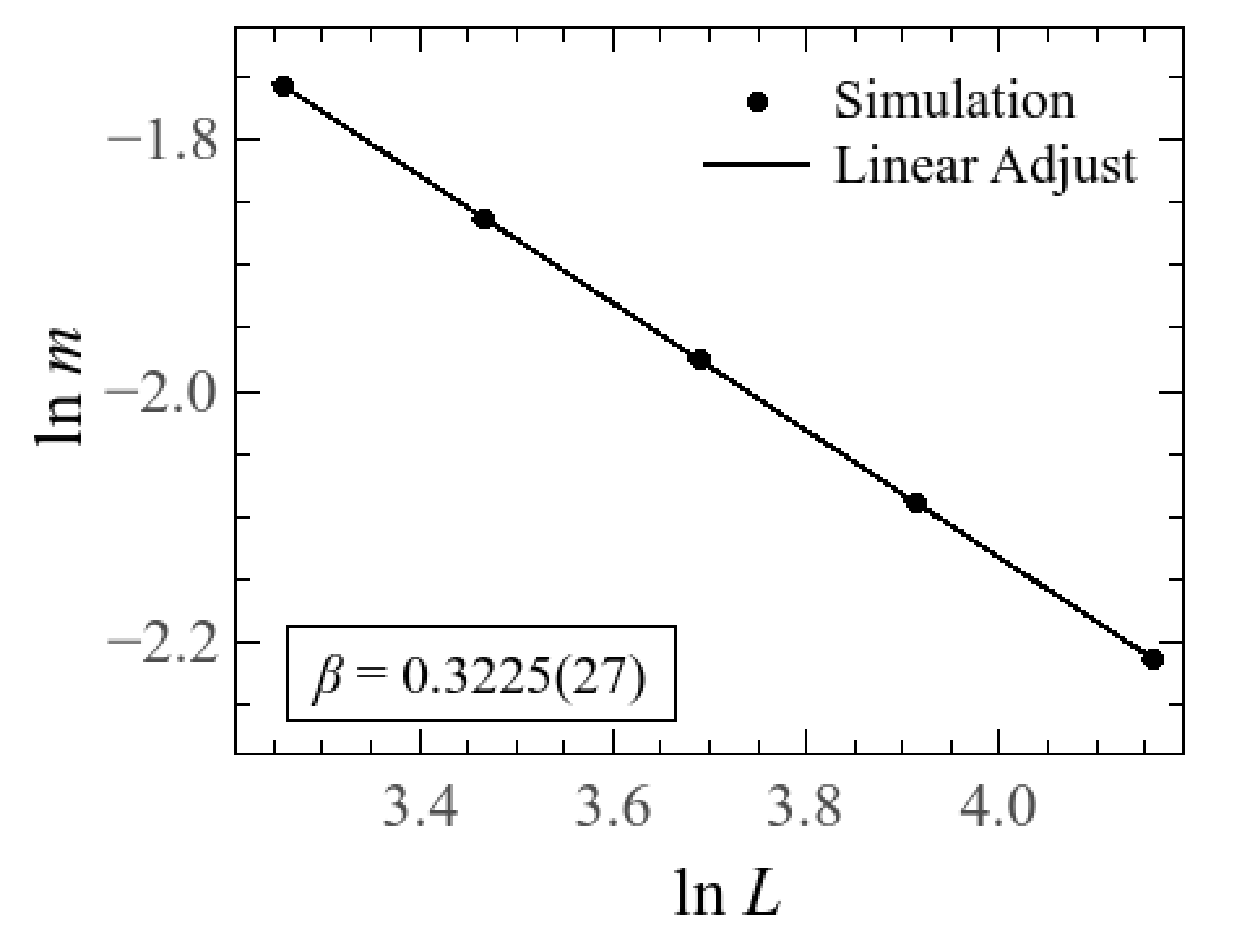}
}
\end{minipage}
\begin{minipage}[t]{0.5\textwidth}\subfigure[]{
\includegraphics[angle=0,width=\textwidth]{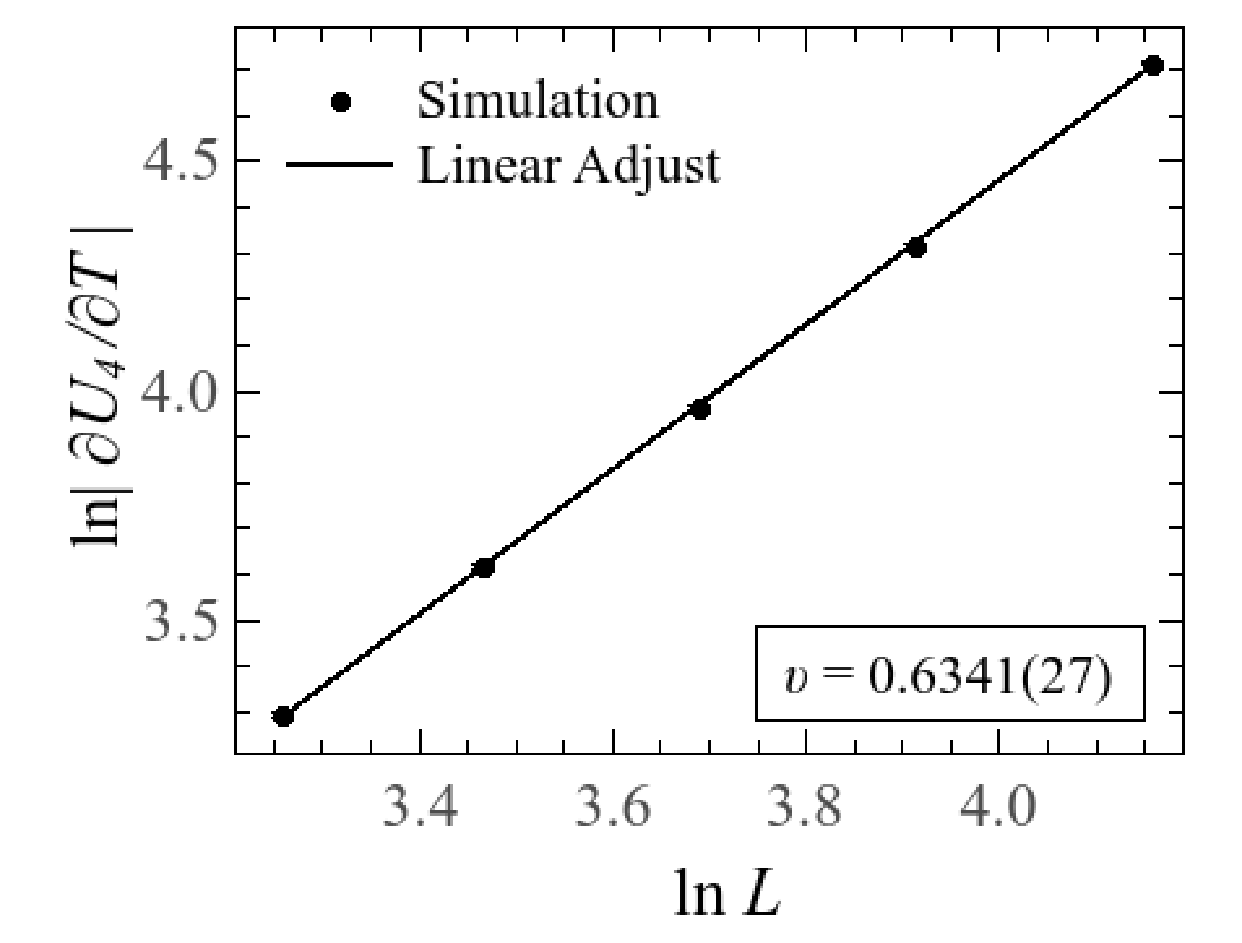}
}
\end{minipage}
\caption{Graphs of the critical exponents $\beta$ and $\nu$ for: (a) and (b) Kagom\'e, (c) and (d) pyrochlore.}
\label{fig:exponents}  
\end{figure}

In the presente work, the simulation on the Kagom\'e lattice consisted of 8 bins with $3 \times 10^6$ MCS each for all the $L = 90$, $100$, $116$, $140$, $166$ and $200$ system sizes.
The behaviour of thermodynamic quantities around the performed temperature is evaluated through the Histogram Technique. The distribution of the $15 \times 8^2 = 960$ Binder Cummulant intersection points of different system sizes shown in Figure \ref{fig:TC} (a) is used to evaluate the critical temperature $T_C$.
On the other hand the critical exponents $\beta$ and $\nu$ are calculated by the size dependence of the magnetization per site $m$ and the derivative of the Binder Cumulant with respect to the temperature ${\partial U_4}/{\partial T}$ given by Equation \ref{eq:exponents} and it is shown in Figure \ref{fig:exponents} (a) and (b).
The results for the pirochlore lattice are obtained in the same way but with 8 bins of size $5 \times 10^6$ with system sizes $L = 26$, $32$, $40$, $50$ and $64$.
The distribution of the $10 \times 8^2 = 640$ values of $T_C$ is in Figure \ref{fig:TC} (a) while the critical exponents determination is in Figure \ref{fig:exponents} (c) and (d).
In all graphs the error bars are smaller than the dots.

The critical temperature and exponents of the Kagom\'e lattice are shown in Table~\ref{tab:Results_Kagome} besides the exact value of $T_C$~\cite{Syozi1951} and the critical exponents of the square lattice~\cite{Stanley1971} which must be the same for Kagom\'e~\cite{Garcia-Adeva2001}.
The results are very accurate taken into account the error bars.
For the pyrochlore lattice the critical temperature and exponents are shown in Table~\ref{tab:Results_pyrochlore} next to the reference value of $T_C$ obtained by Effective Field Theory~\cite{Garcia-Adeva2001} and the critical exponent for the cubic lattice~\cite{Folk2003}.
The critical temperature agrees with expectations taking into account the Kagom\'e lattice results and the distinct values compared to Effective Field Theory are expected as the result of their own approaches.
The critical exponents are in concordance with reference values.
Given the simplicity of this method compared to the common representation it can be used as an alternative to simulations on the Kagom\'e and pyrochlore lattices.
Although spins in these materials found in nature cannot be treated collinearly, this method can be used as a prototype for more complex cases.

\begin{table}[!h]
\begin{minipage}[b]{0.45\linewidth}\centering
  \caption{Critical temperature and exponents for the Kagom\'e lattice.}
  \label{tab:Results_Kagome}
  \begin{tabular}{ccc}
    \hline
       ~    & \multicolumn{2}{c}{Kagom\'e} \\
    \hline       
       ~    & Exact Value  & This Work \\
    \hline    
    $T_C$   & $2.14331944$ & $2.143313(59)$ \\
    $\beta$ & $0.1250$     & $0.1251(11)$   \\
    $\nu$   & $1$          & $0.9989(71)$   \\
    \hline
  \end{tabular}
\end{minipage}
\hspace{0.5cm}
\begin{minipage}[b]{0.45\linewidth}
\centering
  \caption{Critical temperature and exponents for the pyrochlore lattice.}
  \label{tab:Results_pyrochlore}
  \begin{tabular}{ccc}
    \hline
       ~    & \multicolumn{2}{c}{Pyrochlore}   \\
    \hline       
       ~    & Reference Value & This Work      \\
    \hline    
    $T_C$   & $4.347826$      & $4.213893(31)$ \\
    $\beta$ & $0.3258(14)$    & $0.3225(27)$   \\
    $\nu$   & $0.6304(13)$    & $0.6341(27)$   \\
    \hline
  \end{tabular}
\end{minipage}
\end{table}

\section*{Acknowledgments}
A. L. Passos  is grateful for the partial support provided by FAPITEC/SE.




\end{document}